# Strong absorption and selective thermal emission from a mid-infrared metamaterial

J. A. Mason, S. Smith, and D. Wasserman[*]

*Department of Physics ,University of Massachusetts Lowell, One University Avenue, Lowell, MA, 01854, USA*

[*]*Corresponding author: daniel_wasserman@uml.edu*

We demonstrate thin-film metamaterials with resonances in the mid-infrared wavelength range. Our structures are numerically modeled and experimentally characterized by reflection and angularly-resolved thermal emission spectroscopy. We demonstrate strong and controllable absorption resonances across the mid-infrared wavelength range. In addition, the polarized thermal emission from these samples is shown to be highly selective and largely independent of emission angles from normal to 45 degrees. Experimental results are compared to numerical models with excellent agreement. Such structures hold promise for large-area, low-cost metamaterial coatings for control of gray- or black-body thermal signatures, as well as for possible mid-IR sensing applications.

The emergence of the field of metamaterials has led to the development of artificial materials with pre-defined optical properties not obtainable in nature. The potential applications for metamaterials are numerous, with the most commonly cited being the development of optical cloaks [1, 2] and superlenses [3-6]. In the mid-infrared (mid-IR) wavelength range (2-30μm), where all room temperature biological materials and mechanical objects emit thermal radiation, this ability to tailor the optical properties of materials has potentially significant implications. In particular, the development of metamaterial coatings for grey- or black-body emitters would allow for the alteration of the emitters' thermal signatures, 'fooling' thermal sensors designed to detect the coated objects.

In this work we demonstrate selective thermal emission from 1D metamaterial films with controllable resonances in the mid-IR spectral range. The metamaterials demonstrated here are easily fabricated and can be removed from their carrier substrate, giving a thin-film mid-IR metamaterial wall-paper for coating thermal emitters. Our structures are characterized by reflectivity and angularly-dependent emissivity measurements, and show a substantial modification of the thermal characteristics of the bulk materials coated. In addition, we model our structures numerically, and good agreement with our experimental results is observed.

Fundamentally, the basis for selective thermal emission lies the ability to control a material's absorbtivity, $\alpha(\lambda)$, which, by Kirchoff's Law, is equivalent to controlling the material emissivity, $\varepsilon(\lambda)$ [7]. For any material, then, we can experimentally determine emissivity by accounting for all of the light not absorbed by the material (reflected, scattered, and transmitted). A material with subwavelength surface roughness and geometry can be treated as a homogeneous material, thus removing the scattering term [8]. With the inclusion of an optically thick layer of metal as the material ground plane (preventing transmission), Kirchoff's Law reduces to the simplified $\alpha(\lambda)=\varepsilon(\lambda)=1-r(\lambda)$, where $r(\lambda)$ is the wavelength-dependent reflectivity.

Perfect absorbers and selective emitters based on plasmonic and phononic metamaterials have been demonstrated recently at IR frequencies [9-12]. Similar structures have also been proposed and numerically and analytically characterized at THz and near-IR frequencies [13, 14]. Our structures are quite similar to those described in the references above [9, 13], and consist of a three layer system: an optically thick layer of gold (Au) below a layer of spin-on-glass (SOG) into which is integrated a periodic array of Au strips. While earlier work utilized circular plasmonic 'pucks', giving a polarization insensitive optical response, our 1D stripes will result in strong absorption signatures for only TM-polarized light (electric field perpendicular to the stripe direction, as shown in Fig. 1). Despite this, the underlying physics of the structures remains similar. The distinctive optical characteristics of our structures can be qualitatively understood as follows: for TM-polarized light, the stripes act as optical antennas, coupling in incident radiation with a resonance based upon the stripe width (as well as the surrounding dielectric). The incident radiation couples strongly to the magnetic moment created by the anti-parallel surface currents in the stripe and the metal ground plane. The confinement of electromagnetic energy in the dielectric spacer between the stripes and the ground plane results, ideally, for a dielectric spacer of the correct thickness, in 100% absorption. In Ref. [9], the absorption of energy occurs in the gold layers. For our structures, as we will show, absorption occurs in both the dielectric spacer and the Au layers.

A schematic of our sample structure, as well as a scanning electron micrograph of a representative sample is shown in Fig. 1. The structures are fabricated by first patterning the 50nm-thick Au stripes with periodicity $\Lambda=4.8\mu m$ onto a sacrificial substrate. The SOG (Honeywell Accuglass 211) is then spun (thickness t=250nm) above the Au stripes and cured, and subsequently coated with the thick (200nm) Au layer. The sacrificial substrate is removed and the thin film attached to a glass slide, solid Au side down. Samples with stripe widths of w=1.9, 2.26, and 3.18μm were fabricated and characterized.

Samples were first studied by reflection spectroscopy. Here, normally incident broadband mid-IR light from a Bruker V70 Fourier Transform Infrared (FTIR) spectrometer was focused onto the surface of the patterned material through a mid-IR beam splitter. Light reflected from the sample surface was filtered through a polarizer and collected by an external HgCdTe (MCT) detector. The reflection signal was normalized to the polarized reflection from a flat gold mirror and the reflectivity spectrum was calculated. By Kirchoff's Law, dips in the reflection spectra, for wavelengths greater than the stripe periodicity, indicate coupling to absorption resonances.

Our samples were subsequently characterized by thermal emission spectroscopy. Here the patterned sample was heated to 160°C using a custom-built hot-plate mounted on a rotational stage. The emission from the sample was spatially filtered through a slit parallel to the axis of rotation (to minimize the angular spread of the collected light), collected with a 2 in. diameter, 5 in. focal length ZnSe lens, and analyzed with the FTIR spectrometer. A polarizer between the collecting lens and the FTIR was used to select the polarization of the emitted light. The collected spectra were compared to a calibrated blackbody source (Infrared Industries model 101C) at 160°C. The background thermal emission of the experimental set-up (without sample or heaters) was collected and subtracted from each spectrum (sample and calibrated blackbody). The background-corrected emission spectra from our samples are then normalized to the background-corrected blackbody source.

For the angular dependence of the thermal emission, the sample was rotated about the focal point of the collection lens, with emission spectra collected in 10° rotational increments. Angularly-dependent spectra were measured for TM polarized light for rotations in both φ and θ (as shown in Fig. 1(a)), and background-corrected and normalized to the calibrated blackbody.

Our samples showed strong absorption resonances at wavelengths determined by the device stripe width, and not stripe periodicity, indicating that the observed resonances are localized to individual stripes (Fig. 2(b)). For the device with 3.18μm stripe widths, 100% absorption was achieved at a wavelength of 8 μm. In order to understand the nature of these absorption resonances, the optical properties of the spin-on-glass layer were obtained by broad-band ellipsometry measurements, shown in Fig. 2(c). The SOG exhibits a number of absorption peaks, with the dominant peak at $\lambda \approx 9 \mu m$. A weaker SOG absorption is also seen at $\lambda \approx 7.85 \mu m$. For device geometries giving absorption resonances far from the 7.85μm SOG absorption line, this feature appears as a dip in the normal incidence reflection spectra. However, for the w=3.18μm stripes, where the geometric absorption resonance is well-aligned with the 7.85μm material absorption line, a clear splitting is observed in the device reflection spectra giving a local increase in the device reflectivity at the material resonance, and suggesting a coupling between our device's geometric resonance and the material resonance in the SOG.

Our structures were numerically simulated with commercial FEM software [15], using the experimentally determined SOG complex index of refraction. Reflectivity plots for our numerical models are shown in Fig. 2(a). The simulated data shows excellent agreement with our experimental results, including the reproduction of the effects of the material absorption resonance on the reflectivity from the structures.

The demonstration of strong absorption in our samples should result in strongly selective emission at the absorption peaks upon sample heating. Fig. 3 shows the background corrected and normalized emission spectrum from the w=3.18μm structure, as well as the reflectivity from the same sample, for comparison. As expected, spectral dips in the sample's reflection spectrum are mirrored by peaks in the selective emission of the sample. It is important to note that although the emissivity rises above unity at resonance, we do not claim that our samples will emit more power, at certain wavelengths, than a blackbody at equivalent temperature. Such an anomaly has been noted in previous selective emission work [16], and was eventually determined to be a result of localized heating of the selective emitter beyond the bulk material temperature [17]. As radiative emission away from the sample resonance is suppressed, the quenching of these radiative channels results in a localized heating of the thin metamaterials membrane. Because we measure sample temperature on the carrier substrate, and not the film itself, the substrate temperature is expected to underestimate the actual local temperature of the metamaterial film.

Finally, angular emission from the surface of the heated sample (w=3.18μm) was collected for emission angles between 0 and 45° in both the θ and φ directions (see Fig. 1(a)) in order to determine the angular dependence of thermal emission from our structures (Fig 4). In addition, the angular dependence of the sample reflectivity was numerically modeled for varying θ, and the resulting 1- r(λ) plotted for angles θ=0-45°. A strong selective emission is seen at λ~8 μm, and its insensitivity to angle is in good agreement with both our numerical simulations and the results of ref. 9. Slight spectral shifts in the selective emission peak can be seen for angled emission at higher φ, an effect which has also been noted in previous work.

In summary, we have designed, fabricated, and characterized thin-film mid-IR metamaterials with strong absorption resonances. The fabricated structures, as expected, demonstrate strongly selective polarized thermal emission at the designed wavelength. Our experimental results show excellent agreement to numerical simulations. In addition, we qualitatively determine the absorption/emission signals to be a result of both geometrical and material resonances. The development of thicker mid-IR transparent dielectric spacer layers will allow for structures with geometry-based perfect absorption resonances. Conversely, the strong spectral modulation from weak absorption resonances in the dielectric spacer layer suggests the potential of similar structures for mid-IR sensing applications. In total, the device geometries presented show significant potential for the further development of mid-IR metamaterial-based devices and thin films for sensing and security applications.

This work was supported by the National Science Foundation (Award #: 0925542 and the UMass Lowell GK-12 Waves and Vibes program). The authors would also like to thank James Ginn of Sandia National Labs for mid-IR ellipsometry measurements on the spin-on-glass used in this work.

**References.**

1. J. B. Pendry, D. Schurig, and D. R. Smith, Science 312, 1780 (2006).
2. D. Schurig, J. J. Mock, B. J. Justice, S. A. Cummer, J. B. Pendry, A. F. Starr, and D. R. Smith, Science 314, 5801 (2006).
3. V. G. Veselago, Sov. Phys. Usp. 10, 509 (1968).
4. J. B. Pendry, Phys. Rev. Lett. 85, 3966 (2000).
5. A. Grbic and G.V. Eleftheriades, Phys. Rev. Lett., 92, 117403 (2004
6. N. Fang, H. Lee, C. Sun and X. Zhang, Science 308, 534 (2005).
7. J.J. Greffet and M. Nieto-Vesperinas, J. Opt. soc. Am. A 10, 2735 (1998).
8. S. M. Rytov, Sov. Phys. JETP 2, 466 (1955).
9. N. Liu, M. Mesch, T Weiss, M. Hentschel, and H. Giessen, Nano Lett., 10, 2342 (2010).
10. S. O. C. Giraud and D. G. Hasko, Microelectron. Eng. 41/42, 579 (1998).
11. J.-J. Greffet, R. Carminati, K. Joulain, J.-P. Mulet, S. Mainguy, and Y. Chen, Nature (London) 416, 61−64 (2002).
12. J. A. Mason, D. C. Adams, Z. Johnson, S. Smith, A. W. Davis, and D. Wasserman, Opt. Express 18, 25192 (2010).
13. M. Diem, T. Koschny, and C. M. Soukoulis, Phys. Rev. B 79, 033101 (2009).
14. Y. Avitzour, Y. A. Urzhumov, and G. Shvets, Phys. Rev. B 79, 045131 (2009).
15. www.comsol.com
16. S. Y. Lin, J. G. Fleming, and J. Moreno, Appl. Phys. Lett. 83, 380 (2003).
17. J. G. Fleming, Appl. Phys. Lett. **86**, 249902 (2005).

**Figure Captions.**

**Figure 1:** (a) Schematic of the mid-IR metamaterial absorber and (b) scanning electron micrograph of the sample.

**Figure 2:** (a) Simulated and (b) experimental reflectivity from metamaterial absorbers for stripe widths of 1.9 µm (blue), 2.26 µm (green) and 3.18 µm (red). (c) Experimentally determined real (black) and imaginary (orange) components of the spin-on-glass refractive index.

**Figure 3:** Reflectivity (blue) and emissivity (red) of metamaterial absorber structures with stripe widths of 3.18 µm. Emission data was taken at 160°C and normalized to a calibrated blackbody source at the same temperature.

**Figure 4:** Angular dependence of thermal emission from 3.18 µm stripe width metamaterial sample for rotations in (a) θ (experimental), (b) φ (experimental) and (c) θ (numerical). (d) Schematic of experimental set-ups for data collected in (a) and (b).

**Fig. 1**

(a)

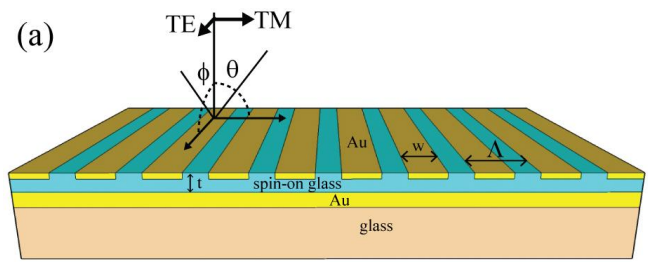

(b)

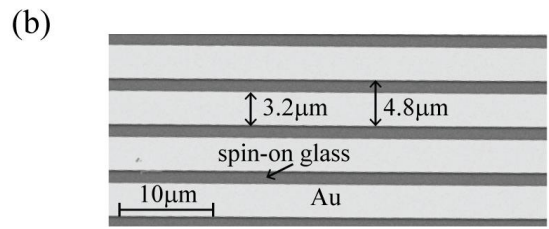

**Fig. 2**

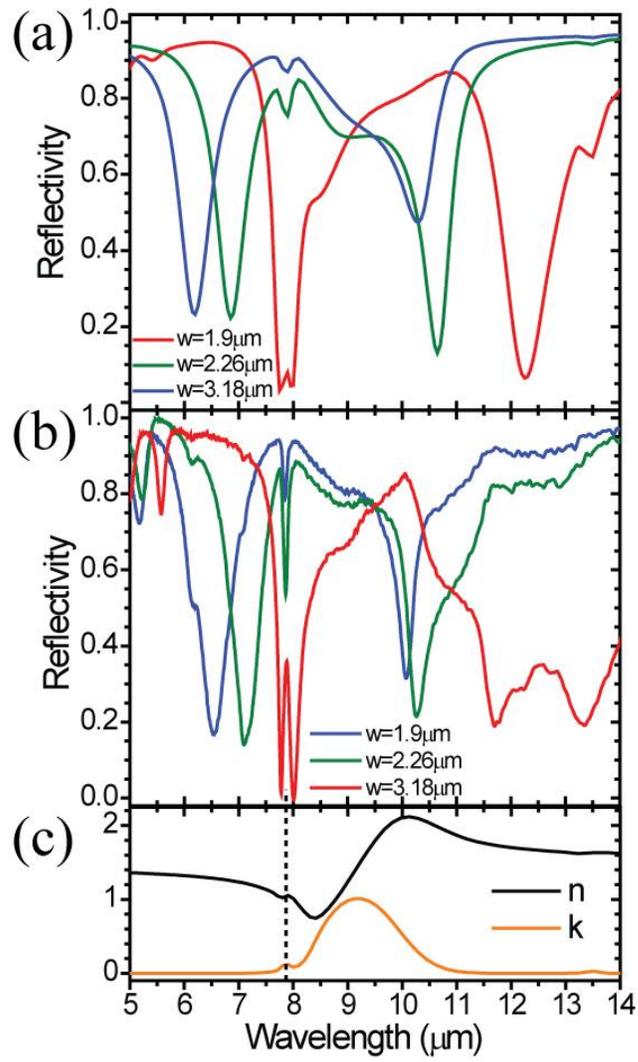

**Fig. 3**

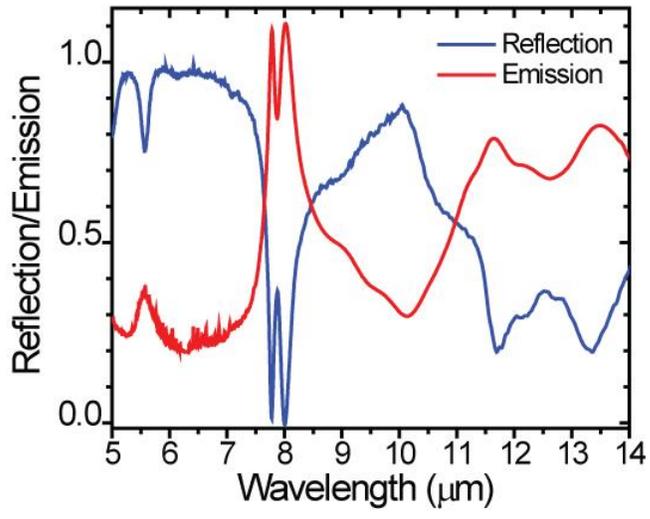

**Fig. 4**

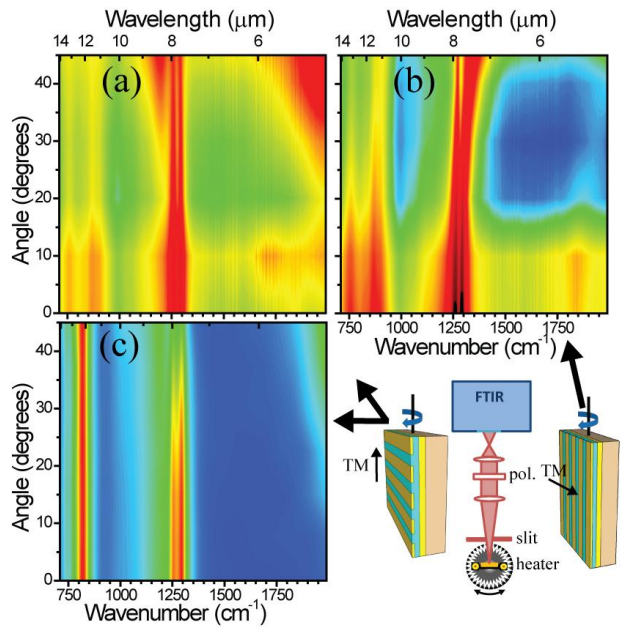